\documentclass
[superscriptaddress,secnumarabic,amssymb,amsmath,nobibnotes,aps,prd,showkeys,showpacs,nofootinbib,onecolumn,notitlepage,noshowpacs]{revtex4}%
\usepackage{setspace}
\usepackage{xcolor}
\usepackage{amsmath}
\usepackage{amsfonts}
\usepackage{verbatim}
\usepackage{amssymb}
\usepackage{graphicx,bm}
\usepackage{graphicx}
\usepackage{bbm}
\usepackage{amsmath}
\usepackage{amssymb}
\usepackage{bbm}
\usepackage{amssymb}
\usepackage{graphicx,bm}
\usepackage{graphicx}
\usepackage{bbm}
\usepackage{epstopdf}
\usepackage[caption=false]{subfig}%
\usepackage{soul}
\providecommand{\U}[1]{\protect\rule{.1in}{.1in}}

\newcommand{\be}{\begin{equation}}
\newcommand{\ee}{\end{equation}}

\newcommand{\mincir}{\raise
-3.truept\hbox{\rlap{\hbox{$\sim$}}\raise4.truept\hbox{$<$}\ }}
\newcommand{\magcir}{\raise
-3.truept\hbox{\rlap{\hbox{$\sim$}}\raise4.truept\hbox{$>$}\ }}

\ifx\pdfoutput\relax\let\pdfoutput=\undefined\fi
\newcount\msipdfoutput
\ifx\pdfoutput\undefined\else
\ifcase\pdfoutput\else
\ifx\paperwidth\undefined\else
\ifdim\paperheight=0pt\relax\else\pdfpageheight\paperheight\fi
\ifdim\paperwidth=0pt\relax\else\pdfpagewidth\paperwidth\fi
\fi\fi\fi
\begin{document}
\title{Quantization of Einstein-aether Scalar field Cosmology}
\author{N. Dimakis}
\email{nsdimakis@scu.edu.cn; nsdimakis@gmail.com}
\affiliation{Center for Theoretical Physics, College of Physics,
Sichuan University, Chengdu 610064, China}
\author{T. Pailas}
\email{teopailas879@hotmail.com}
\affiliation{Nuclear and Particle Physics section, Physics Department, University of
Athens, 15771 Athens, Greece}
\author{A. Paliathanasis}
\email{anpaliat@phys.uoa.gr}
\affiliation{Institute of Systems Science, Durban University of Technology, Durban 4000,
South Africa}
\author{G. Leon}
\email{genly.leon@ucn.cl}
\affiliation{Departamento de Matem\'{a}ticas, Universidad Cat\'{o}lica del Norte, Avda.
Angamos 0610, Casilla 1280 Antofagasta, Chile.}
\author{Petros A. Terzis}
\email{pterzis@phys.uoa.gr}
\affiliation{Nuclear and Particle Physics section, Physics Department, University of
Athens, 15771 Athens, Greece}
\author{T. Christodoulakis}
\email{tchris@phys.uoa.gr}
\affiliation{Nuclear and Particle Physics section, Physics Department, University of
Athens, 15771 Athens, Greece}

\begin{abstract}
We present, for the first time, the quantization process for the
Einstein-aether scalar field cosmology. We consider a cosmological theory
proposed as a Lorentz violating inflationary model, where the aether and
scalar fields interact through the assumption that the aether action constants
are ultra-local functions of the scalar field. For this specific theory there
is a valid minisuperspace description which we use to quantize. For a
particular relation between the two free functions entering the reduced
Lagrangian the solution to the Wheeler-DeWitt equation as also the generic
classical solution are presented for any given arbitrary potential function.
\end{abstract}
\maketitle
\date{\today}

\section{Introduction}

\label{sec1}

The interest in Lorentz violating cosmological theories comes together with
the idea of a variable speed of light. The Horava-Lifshitz gravity is
certainly violating Lorentz invariance by construction, since arbitrary time
re-definitions are not among its covariances \cite{hor-lif}. In the
Einstein-aether theory, a unit time-like rotationally invariant vector field,
called the ``aether" is also responsible for the same effect \cite{jacob2004}.
Specifically, in Einstein-aether theory quadratic kinematic quantities of the
unitary time-like vector fields are introduced in the gravitational Action
Integral \cite{Carru,jac01}. These new terms break the Lorentz symmetry
\cite{esf}, by selecting a preferred frame at each space time point,
while keeping the field equations of second-order as in the case of General
Relativity.

On the other hand, scalar fields play a prominent role in modern cosmology.
The main mechanism for the description of the inflation is based on the
domination of a scalar field potential, known as inflaton \cite{guth}.
Moreover, scalar fields have also been proposed as dark energy models, while
they can attribute the geometrodynamical degrees of freedom provided by
higher-order theories which belong to the class of modified theories of
gravity, for more details we refer the reader to
\cite{Ratra,Barrow,Linder,Copeland,Overduin,cl1,cl2,sf1,sf2,sf3,sf4,sf5,sf6,sf7,sf8,sf9}; for a complete review on the cosmological implications in modified theories of gravity see \cite{Ishak} and references therein.

One can contemplate a non-trivial coupling of the scalar to the aether field by allowing the
coefficients of its kinetic action to be functions of the scalar field
\cite{carroll2004}. Such a theory has been proposed before as an alternative
inflationary model which provides two periods of inflation \cite{Kanno:2006ty}. A classical slow-roll era and a Lorentz violating epoch. In this work we
realize this idea for the case of a spatially flat FLRW universe and a scalar field with
arbitrary potential and we present, for the first time, the quantization of
the Einstein-aether scalar field cosmology; we also derive the generic
algebraic classical solution to the field equations. Einstein-aether scalar field theory are of special interest in the
scientific society and there are various studies in the literature on the
subject, some of these studies for homogeneous and inhomogeneous spacetimes
can be found in
\cite{Coley:2015qqa,Latta:2016jix,Alhulaimi:2017ocb,VanDenHoogen:2018anx,Coley:2019tyx,Leon:2019jnu,Paliathanasis:2020bgs,Paliathanasis:2019pcl}%
.

The structure of the paper is as follows. In Section \ref{sec2}, we define the cosmological model that we focus on in
this work; it is the Einstein-aether scalar field cosmology in a homogeneous
and isotropic geometric background space where only quadratic terms of the
derivatives exist in the Action Integral. The latter property is essential for
enabling a minisuperspace description of the gravitational field equations. In
Section \ref{sec3} we present the point-like Lagrangian of our model. This
specific model has been proposed before as an alternative model for inflation.
The quantization process is presented in Section \ref{sec5}. In Section
\ref{sec6} we study the classical limit while we discuss our results and we
draw our conclusions in Section \ref{sec7}.

\section{Einstein-aether scalar field cosmology}

\label{sec2}

Kanno and Soda in \cite{Kanno:2006ty} proposed a Lorentz violating
Einstein-aether Action Integral while assuming the Einstein-aether coupling
parameters to be functions of this scalar field, thus generating a coupling
between the scalar and the aether field. The main characteristic of this model
is that the inflationary epoch can be described by two stages; the usual
slow-roll stage and the new Lorentz violating stage.

A more general Einstein-aether scalar field model was later proposed in
\cite{DJ}, where a scalar field is introduced in the Einstein-aether Action
Integral with the scalar field potential being a function of the field and the
kinematic invariants of the aether field. The model of Kanno and Soda is
included as a special case in that of \cite{DJ}. While\ the model proposed in
\cite{DJ} describes a second-order theory, only in the limit of
\cite{Kanno:2006ty} the Action Integral depends only on quadratic terms of the
derivatives. As we shall see in the following this is an essential property in
order for the dynamical field equations to admit a minisuperspace description.

The Einstein-aether scalar field model proposed in \cite{Kanno:2006ty} is
described by the Action Integral%
\begin{equation}
S=\int d^{4}x\sqrt{-g}\left(  \frac{R}{2}-\frac{1}{2}g^{\mu\nu}\phi_{;\mu}%
\phi_{;\nu}-V\left(  \phi\right)  \right)  -S_{Aether}, \label{ac.01}%
\end{equation}
where $S_{Aether}$ describes the terms of the aether field $u^{\mu}$ as
follows
\begin{equation}%
\begin{split}
S_{Aether}= \int dx^{4}\sqrt{-g}\mathcal{L}_{Aether} =\int d^{4}x\sqrt{-g}\Big[  &  \beta_{1}\left(  \phi\right)
u^{\nu;\mu}u_{\nu;\mu}+ \beta_{2}\left(  \phi\right)  \left(  g^{\mu\nu}%
u_{\mu;\nu}\right)  ^{2} + \beta_{3}\left(  \phi\right)  u^{\nu;\mu}u_{\mu
;\nu}\\
&  +\beta_{4}\left(  \phi\right)  u^{\mu} u^{\nu}u_{\kappa;\mu}u^{\kappa}_{\; ;\nu}%
-\lambda\left(  u^{\mu}u_{\nu}+1\right)  \Big].\label{ac.02}%
\end{split}
\end{equation}

Function $\lambda$ is the Lagrange multiplier which is introduced to ensure
the unitarity of the aether field, i.e. $u^{\mu}u_{\mu}+1=0$. Coefficients
$\beta_{1},~\beta_{2},~\beta_{3}~$and $\beta_{4}$ define the coupling between
the aether and the gravitational field. While in the Einstein-aether theory
the coefficients are constants in this specific theory they are functions of
the scalar field $\phi$.

We select the case of a homogeneous and isotropic spacetime described by the
spatially flat FLRW line element%
\begin{equation}
ds^{2}=-N^{2}\left(  t\right)  dt^{2}+a^{2}\left(  t\right)  \left(
dx^{2}+dy^{2}+dz^{2}\right)  , \label{ac.03}%
\end{equation}
where $a\left(  t\right)  $ is the scale factor, $N\left(  t\right)  $ is the
lapse function. The corresponding Hubble function is defined as $H\left(
t\right)  =\frac{1}{N}\frac{\dot{a}}{a}$ (where a dot denotes total derivative
with respect to the variable $t$) \ and for the aether field we have $u^{\mu
}=\frac{1}{N}\delta_{t}^{\mu}$. These assumptions imply that $\phi=\phi(t)$
and then the gravitational field equations follow from the variation of the
action \cite{Kanno:2006ty}
\begin{equation}
-\frac{3}{N^{2}}B\left(  \phi\right)  a\dot{a}^{2}+\frac{1}{2N^{2}}a^{3}%
\dot{\phi}^{2}+a^{3}V\left(  \phi\right)  =0. \label{ac.08}%
\end{equation}%
\begin{equation}
2a\left(  \ddot{a}-\frac{1}{N}\dot{a}\dot{N}\right)  B\left(  \phi\right)
+2aB_{,\phi}\dot{a}\dot{\phi}+B\left(  \phi\right)  \dot{a}^{2}+\frac{1}%
{2}a^{2}\dot{\phi}^{2}-N^{2}a^{2}V\left(  \phi\right)  =0, \label{ac.06}%
\end{equation}%
\begin{equation}
\ddot{\phi}+3\frac{\dot{a}}{a}\dot{\phi}-\frac{\dot{N}}{N}\dot{\phi}+\frac
{3}{a^{2}}B_{,\phi}\dot{a}^{2}+N^{2}V_{,\phi}=0. \label{ac.07}%
\end{equation}
where the new function $B\left(  \phi\right)  $ is expressed as $B\left(
\phi\right)  =\beta_{1}\left(  \phi\right)  +3\beta_{2}\left(  \phi\right)
+\beta_{3}\left(  \phi\right)  +1$. \ We observe that in the limit where
$B\left(  \phi\right)  =$const., the field equations take the form of the
quintessence scalar field model in General Relativity, which means that the
Lorentz violating inflationary stage does not exist. Thus in the following we
consider the case where $B\left(  \phi\right)  _{,\phi}\neq0.$

In order to demonstrate the correspondence with the General Relativistic limit we take the Einstein equations that emerge for the spacetime \eqref{ac.03} in presence of a perfect fluid source. The set of equations $R_{\mu\nu}-\frac{1}{2}g_{\mu\nu}=\kappa T_{\mu\nu}$, where $T^\mu_{\;\; \nu}=\mathrm{diag}(-\rho,p,p,p)$ is the energy momentum of the fluid with energy density $\rho$, pressure $p$ and $\kappa$ the constant related to the gravitational coupling, is equivalent to
\begin{subequations} \label{Fried1}
\begin{align}
  3 H^2 & = \kappa \rho  \\
  -3 H^2 -\frac{2}{N} \dot{H} & = \kappa p.
\end{align}
\end{subequations}
In the latter we have used the Hubble function $H=\frac{\dot{a}}{N a}$ as expressed in an arbitrary time gauge. In the cosmic time gauge, where $N=1$ the above equations fall to the usual expressions given in the literature for the Friedmann equations \cite{Weinberg}.

In a similar spirit we may start from the cosmological field equations (\ref{ac.08})-(\ref{ac.06}) and write
\begin{subequations}\label{Fried2}
\begin{align}
3H^{2} & =k_{eff}\rho_{eff}\\
-\left(  2\dot{H}+3H^{2}\right)  & =k_{eff}p_{eff}%
\end{align}
\end{subequations}
where the new functions~$\rho_{eff}$ and $p_{eff}$ are the energy density and
pressure for the effective fluid defined as $\rho_{eff}=\frac{1}{2N^2}\dot{\phi
}^{2}+V\left(  \phi\right)  $ , $p_{eff}=\left(  \frac{2B_{,\phi}H}{N}\dot{\phi}%
+\frac{1}{2N^2}\dot{\phi}^{2}-V\left(  \phi\right)  \right)  $. Additionally,
$k_{eff}= B\left(  \phi\right)  ^{-1}$ is not a constant but varies in
time, that is an effect that is observed in the Jordan frame and in scalar
tensor theories, however here this effect follows from the time-dependent
coupling function for the aether field. We can see thus how the $B(\phi)=$const. limit turns equations \eqref{Fried2} into \eqref{Fried1} as $k_{eff}=\kappa$ and $\rho=\rho_{eff}$, $p=(p_{eff})|_{B'(\phi)=0}$ with the $\rho$ and $p$ corresponding now to the typical energy density and pressure that is produced by a minimally coupled scalar field to a FLRW spacetime. In this context, the $\rho_{eff}$ and $p_{eff}$ of the full equations, when $B(\phi)\neq$constant can be seen as the energy density and pressure of a fluid generated by a non-minimally coupled scalar field.

\section{Minisuperspace description}

\label{sec3}

The superspace is an infinite-dimensional space serving as the basic
configuration space of canonical quantum gravity \cite{DeWitt}. As defined in the canonical
formulation of General Relativity, it consists of all Riemannian 3-dimensional
metrics and the matter fields. In cosmology, due to the spacetime symmetries
of the geometry, the infinite degrees of freedom of the corresponding
superspace are truncated to a finite number and thus a particular
minisuperspace model is achieved. For example, consider the typical cosmological line element
\begin{equation}
  ds^2 = - N(t)^2 dt^2 + \gamma_{AB}(t) \sigma^A_i(x) \sigma^B_j(x) dx^i dx^j, \quad A,B,i,j=1,2,3
\end{equation}
which describes Bianchi cosmologies. The $\gamma_{AB}$ is the scale factor matrix, the $\sigma^A_i$ are the Cartan forms corresponding to the three dimensional group of isometries that characterizes the Bianchi model and carry only dependence on the spatial variables. The reduction of a generic gravitational action incorporating such a model is realized as follows:
\begin{equation}
S=\int d^{4}x\sqrt{-g}\mathcal{L}(g,\phi,u)  = \int d^3x \sigma(x) dt N(t) \sqrt{\gamma} \mathcal{L}(g,\phi,u)  \rightarrow S = \int dt L\left(  N,q^I,\dot
{q}^I\right),
\end{equation}
where $\gamma$, $\sigma$ are the determinants of $\gamma_{AB}$ and $\sigma^A_i$ respectively.  With $(g,\phi,u)$ we denote the possible dependence of a generic Lagrangian on scalars of the metric $g_{\mu\nu}$, the scalar $\phi$ and the vector field $u_\mu$, as well as their derivatives. In cosmology, the spatial dependence can be integrated out of the
action, due to the space-time symmetries, leaving only a multiplicative
constant $\mathcal{V}_{0}=\int\sigma(x) d^3x$ symbolizing the volume of a finite three-space
cell. The only dynamical part that remains in this reduction is $t$ dependent and expressed by the point Lagrangian $L\left(N,q^I,\dot{q}^I\right)$, where the $q^I(t)$ are the remaining $t$ dependent degrees of freedom after the reduction has taken place. The range of the capital index $I$ counts through the dimension of the minisuperspace and apart from the number of independent $\gamma_{AB}$, it also incorporates any matter degrees of freedom that are still present after the reduction. Of course, one has always to ensure that the variation of the new action of finite degrees of freedom gives rise to Euler-Lagrange equations that are
equivalent to those of the original field theory, under the assumed ansatz for
the metric and the matter fields. In cosmology the $L$ is a singular
Lagrangian given, for matter actions quadratic in the field derivatives, by the
following expression\qquad%
\begin{equation}
L\left(  N,q^I,\dot{q}^I\right)  =\mathcal{V}_{0}\left[  \frac{1}%
{2N}G_{IJ}\left(  q\right)  \dot{q}^I\dot{q}^J-NU\left(
q\right)  \right]  . \label{lan1}%
\end{equation}
For details see for example the reduction performed in Einstein's General Relativity to arrive in a minisuperspace Lagrangian of the form \eqref{lan1} in \cite{MisnerinKlauder}.
Functions $q^I(t),~N(t)~$\ are the unknown functions which describe the
spacetime and the kinematic quantities of the matter source ($N(t)$ correspond
to the lapse-function). The $G_{IJ}\left(  q\right)  $ transforms as a
second-rank tensor under arbitrary redefinitions of the $q^I$s. It is the
so-called minisuperspace metric, while $U\left(  q\right)  $ is the effective
potential which describes the dynamical interactions of the gravitational
field and of the matter source. The Lagrangian function $L$ is a
singular Lagrangian since $\det\left(\frac{\partial^2 L}{\partial\dot{y}^{\tilde{I}}\partial
y^{\tilde{J}}}\right)=0$, where $y^{\tilde{I}}=\left(  N,q^I\right)  $.

Not all the cosmological models in General Relativity have a minisuperspace
description. For a full scale factor matrix and non vanishing shift, only the
Bianchi models which belong to the Class A and the Bianchi V admit a
minisuperspace description (see \cite{MacCallum} and \cite{Sneddon} respectively). Moreover, there are some inhomogeneous models
where the field equations follow from a point-like Lagrangian of the form
(\ref{lan1}). In the context of alternative theories of gravity not all the
proposed theories have a minisuperspace description.

The existence of a Lagrangian function for a given dynamical system, known
also as the inverse problem, is essential in physics. In addition, the
existence of a point-like Lagrangian for the given dynamical system can be
used for the quantization process, which is the main approach applied in
quantum cosmology. From there various approaches can be followed, e.g. canonical theory, loop quantum cosmology, path integrals etc. \cite{Isham,HHwave,Hawking,Vilenkin,Kiefer,Kim,Ashtekar,Bojo}; with interesting implications regarding cosmological effects, e.g. associating the dark energy with the quantum potential \cite{Faraggi}. Due to the lack of a complete theory of Quantum Gravity, quantum cosmology has been the test laboratory to offer at least some hindsight on what we should expect from the full theory. However, we always need to keep in mind that its range of validity highly depends on whether it is correct to neglect the spatial degrees of freedom for these spacetimes of high symmetry which yield minisuperspace models. The situation is not so simple and it has been shown that such reductions may be precarious \cite{Kuchar1}.  The importance of the existence of an equivalent Lagrangian description
of a given set of equations lies in the reach methods of analytical Mechanics
that can be applied in order to study the evolution of the field equations and
their integrability. In the minisuperspace approach, the quantum analogs of
the classical integrals of motion can be used as supplementary conditions in
conjunction to the Wheeler-DeWitt equation, so that the wave function describing quantum states is
defined up to constants.

As far as the Einstein-aether theory is concerned in general, the gravitational
field equations do not admit a point-like Lagrangian. The determination of a
Lagrangian description for the field equations in Einstein-aether theory was
the subject of study in \cite{roum1,roum2}.

For the cosmological model of our consideration, the unknown functions of the
spacetime (\ref{ac.03}) are the scale factor $a$ and the lapse function $N,$
while from the matter source the dynamical variable is function $\phi$. We
observe that the field equations (\ref{ac.08})-(\ref{ac.07}) follow from the
variation of the Action Integral%
\begin{equation}
S=\int dt L\left(  N,a,\dot{a},\phi,\dot{\phi}\right)  , \label{ac.04}%
\end{equation}
where now the Lagrangian function~$L\left(  N,a,\dot{a},\phi,\dot{\phi
}\right)  $ is the point-like Lagrangian \cite{anea1}
\begin{equation}
L\left(  N,a,\dot{a},\phi,\dot{\phi}\right)  =\frac{\mathcal{V}_{0}}{N}\left(
-3B\left(  \phi\right)  a\dot{a}^{2}+\frac{1}{2}a^{3}\dot{\phi}^{2}\right)
-N\mathcal{V}_{0}a^{3}V\left(  \phi\right)  . \label{ac.05}%
\end{equation}
where $q^{I}=\left(  a,\phi\right)$. The minisuperspace metric is%
\begin{equation}
\label{minimetr}G_{IJ}=%
\begin{pmatrix}
-6B\left(  \phi\right)  a & 0\\
0 & a^{3}%
\end{pmatrix}
.
\end{equation}
and the effective potential is $U\left(  q\right)  =a^{3}V\left(  \phi\right)
$.

The derivation of \eqref{ac.04} follows directly by the original action \eqref{ac.01}, which consists of three parts: a) The gravitational
\begin{equation} \label{contr1}
  \sqrt{-g} \frac{R}{2} = -3 \frac{a\dot{a}^2}{N} + \frac{d}{dt}\left( \frac{3 a^2 \dot{a}}{N} \right) ,
\end{equation}
b) the scalar field contribution
\begin{equation}
  -\sqrt{-g}\left(\frac{1}{2}g^{\mu\nu}\phi_{;\mu} \phi_{;\nu}+V\left(  \phi\right) \right) = \frac{a^3 \dot{\phi}^2}{2N} - N a^3 V(\phi)
\end{equation}
and c) that of the Aether
\begin{equation}
   \sqrt{-g} \mathcal{L}_{Aether} =- \frac{3a \dot{a}^2}{N}  \left(\beta_1(\phi)+3\beta_2(\phi)+\beta_3(\phi) \right) =  \frac{3a \dot{a}^2}{N} \left(1-B(\phi)\right)
\end{equation}
where in the last relation the substitution $B(\phi)= 1+\beta_1(\phi)+3\beta_2(\phi)+\beta_3(\phi)$ has been used. By combining the three and ignoring the total derivative appearing in \eqref{contr1} since it gives a surface term in the Lagrangian we arrive at \eqref{ac.05} where a multiplicative constant $\mathcal{V}_{0}=\int d^3 x$ also appears from the finite volume integration of the spatial part of the original action  \eqref{ac.01}.

The metric defined by the kinetic part of the point-like Lagrangian has
dimension two, i.e. $\dim G_{IJ}=2$, which means that it admits an
infinite number of conformal symmetries, independently of the functional form
of $B\left(  \phi\right)  $. Recall that we assume that $B_{,\phi}\left(
\phi\right)  \neq0$.

\section{Quantization}

\label{sec5}

We can exploit the parametrization invariance of Lagrangian \eqref{lan1} to
bring it into an equivalent form which resembles the motion of a free
relativistic particle in a (generally) curved space. To this end, we
reparametrize the lapse function as $N\mapsto n= \frac{2 N a^{3} V(\phi)}{
\mathcal{V}_{0}}$ in \eqref{ac.05} in order to obtain
\begin{equation}
\label{lagn}L \rightarrow L_{n} = \frac{1}{2 n} \bar{G}_{IJ} \dot
{q}^I\dot{q}^J-n \frac{\mathcal{V}_{0}^{2}}{2}.
\end{equation}
Note that $L_{n}$ and $L$ are equivalent, i.e. they reproduce the
same set of Euler-Lagrange equations. Having obtained $L_{n}$ in this form
allows us to interpret $\mathcal{V}_{0}$ as the ``mass" of the supposed
relativistic particle and
\begin{equation}
\bar{G}_{IJ} = 2a^{3} V(\phi) G_{IJ},
\end{equation}
as the scaled mini-superspace metric corresponding to a ``constant (effective) potential" in the Lagrangian, i.e. the metric of
the space in which the motion of the free particle takes place, where $G_{IJ}$ is given by
\eqref{minimetr}. In the particular problem we are studying, $\bar{G}%
_{IJ}$ will generally designate a two dimensional curved manifold of
hyperbolic signature. However, there exists a large class of models for which
this space becomes flat, thus leading to a straightforward quantum description.
Specifically, it is easy to see that, if the potential $V(\phi)$ and the
coupling function $B(\phi)$ are related through
\begin{equation}
\label{lawV}V(\phi) = \frac{V_{1}}{B(\phi)} \exp\left(  V_{2} \int\frac
{1}{\sqrt{B(\phi)}} d\phi\right)  ,
\end{equation}
or equivalently
\begin{equation}
\label{lawB}B(\phi) = \frac{1}{V(\phi)}\left(  B_{1} + B_{2}\int\sqrt{V(\phi)}
d\phi\right)  ^{2},
\end{equation}
where the $V_{i}$, $B_{i}$, $i=1,2$ are constants, then the corresponding
metric $\bar{G}_{IJ}$ is that of a flat space. Relations \eqref{lawV}
and \eqref{lawB} guarantee that the Riemann curvature tensor of the
mini-superspace is zero.

As a result, whenever \eqref{lawV} (or equivalently \eqref{lawB}) holds, the
system is equivalent to a motion of a free relativistic particle in a two
dimensional flat space. Consequently there exist three classical integrals of
motion, whose quantum counterparts can be used as observables in a canonical
quantum description together with the Wheeler-DeWitt equation. Before
proceeding with the quantum description, let us briefly give the connection
with the Cartesian coordinates, say $(u,v)$, with respect to which the
solution of all this class of models can be obtained straightforwardly. The
mini-superspace line element corresponding to the metric $\bar{G}_{IJ}$ is
\begin{equation}
\label{minilel}ds_{2D}^{2} = -12 a^{4} B(\phi)V(\phi) da^{2} + 2 a^{6} V(\phi)
d\phi^{2}.
\end{equation}
By using \eqref{lawB} and introducing a new variable $\phi\mapsto \psi
= \int\!\sqrt{V(\phi)}d\phi$, the expression \eqref{minilel} becomes
$ds_{2D}^{2} = -du^{2} + dv^{2}$ under the transformation
\begin{align}
a  &  = 2^{\frac{15-\sqrt{6} B_{2}}{24 B_{2}^{2}-36}} 3^{\frac{1}{4 \sqrt{6}
B_{2}-12}} \left(  2 B_{2}+\sqrt{6}\right)  ^{\frac{1}{2 \sqrt{6} B_{2}+6}}
\left(  3-\sqrt{6} B_{2}\right)  ^{\frac{1}{6-2 \sqrt{6} B_{2}}}
(u+v)^{\frac{1}{6-2 \sqrt{6} B_{2}}} (u-v)^{\frac{1}{2 \sqrt{6} B_{2}+6}}\\
\psi  &  = -\frac{B_{1}}{B_{2}} +\frac{1}{B_{2}} 2^{\frac{B_{2}
\left(  6 B_{2}+\sqrt{6}\right)  }{12-8 B_{2}^{2}}} 3^{\frac{\sqrt{\frac{3}%
{2}} B_{2}}{6-2 \sqrt{6} B_{2}}} \left(  B_{2}+\sqrt{\frac{3}{2}}\right)
^{\frac{B_{2}}{2 B_{2}+\sqrt{6}}} \left(  3-\sqrt{6} B_{2}\right)
^{\frac{\sqrt{\frac{3}{2}} B_{2}}{\sqrt{6} B_{2}-3}} (u+v)^{\frac{\sqrt
{\frac{3}{2}} B_{2}}{\sqrt{6} B_{2}-3}} (u-v)^{\frac{B_{2}}{2 B_{2}+\sqrt{6}}%
},
\end{align}
for which of course we need to assume $B_{2}\neq0$. In the special case where
$B_{2}=0$ the corresponding transformation is easily derived to be
\begin{equation}
a = \left(  \frac{3 \left(  u^{2}-v^{2}\right)  }{4 B_{1}^{2}} \right)
^{\frac{1}{6}} , \quad \psi = \frac{B_{1}}{\sqrt{6}} \ln\left[  \frac{4
(u+v)}{B_{1}^{2} (u-v)} \right]  .
\end{equation}

We may now proceed with the quantization of the system. The classical
Hamiltonian constraint that emerges from Lagrangian \eqref{lagn} is
\begin{equation}
\mathcal{H} = \frac{1}{2} \bar{G}^{IJ} p_{I}p_{J}+
\frac{\mathcal{V}_{0}^{2}}{2} \approx0
\end{equation}
where $p_{I}=\frac{\partial L_{n}}{\partial\dot{q}^{I}}$ are the
momenta and the symbol ``$\approx$" denotes a weak equality in the Dirac
sense \cite{Dirac1}. In the canonical description we assign to the momenta the differential
operators $p_{\alpha}\rightarrow\hat{p}_{I}= -\mathrm{i} \hbar
\frac{\partial}{\partial q^{I}}$, while for the factor ordering in the
kinetic term of the Hamiltonian constraint we choose to make use of the
Laplacian\footnote{One can also use the conformal Laplacian in more
complicated systems with a higher dimensional non-flat configuration space \cite{tchrisza}.
However here it makes no difference since we have a two dimensional
mini-superspace and there is no distinction between them.} and thus have the
quantum constraint operator
\begin{equation}
\label{Hop}\widehat{\mathcal{H}} = - \hbar^{2} \frac{1}{2} \nabla_{I
}\nabla^{I}+ \frac{\mathcal{V}_{0}^{2}}{2}%
\end{equation}
which - following Dirac's prescription of quantizing constrained systems \cite{Dirac2} - we
demand to annihilate $\Psi$, i.e. $\widehat{\mathcal{H}}\Psi=0$ must hold for
all the states of the system. The latter defines the Wheeler-DeWitt equation
of the mini-superspace model.

In this flat two-dimensional configuration space we are studying, there are
two well known quantization algebras: one involving the constant translations
generators and another the boost in the $u-v$ plane. To utilize the first, we
start from the Cartesian coordinates where $ds^{2}_{2D} = -du^{2}+ dv^{2}$ and
use the two classical integrals of motion, which in these coordinates are just
$p_{u}$ and $p_{v}$. Their quantum counterparts are the commuting operators
$\hat{p}_{u} = -\mathrm{i} \frac{\partial}{\partial u}$ and $\hat{p}_{v} =
-\mathrm{i} \frac{\partial}{\partial v}$ which can be used to define the
eigenvalue equations
\begin{equation}
\hat{p}_{u} \Psi= \mu\Psi, \quad\hat{p}_{v} \Psi= \nu\Psi
\end{equation}
admitting the plane wave solution $\Psi(u,v) = \Psi_{\mu\nu}(u,v) = \frac
{1}{2\pi\hbar}e^{\frac{\mathrm{i}}{\hbar} (\mu u+\nu v)}$, which normalizes to
a product of Dirac delta functions since
\begin{equation}
\int_{-\infty}^{+\infty}\int_{-\infty}^{+\infty} \Psi^{*}_{\mu^{\prime}%
\nu^{\prime}}\Psi_{\mu\nu} du dv = \delta(\mu-\mu^{\prime})\delta(\nu
-\nu^{\prime}) .
\end{equation}
The spectrum is continuous and the quantum numbers $\mu,\nu$ can take values
in the entire $\mathbb{R}$ domain. However, the quantum Hamiltonian constraint
sets the additional condition
\begin{equation}
\widehat{\mathcal{H}}\Psi_{\mu\nu}=0 \Rightarrow\left[  -h^{2} \left(
-\frac{\partial^{2}}{\partial u^{2}}+\frac{\partial^{2}}{\partial v^{2}%
}\right)  +\mathcal{V}_{0}^{2}\right]  \Psi_{\mu\nu} =0 \Rightarrow
\mathcal{V}_{0}^{2} = \mu^{2} - \nu^{2}%
\end{equation}
which forces us to assume that $|\nu| < |\mu|$.

The second way to proceed with the canonical quantization is to use the
quantum equivalent of the third classical integral of motion, which in these
variable is $Q = v p_{u} + u p_{v}$. In this case, it is far more convenient
to utilize coordinates in which the corresponding symmetry generator assumes a
normal form. In particular we may adopt the transformation $u = r \cosh\theta
$, $v = r \sinh\theta$, which makes the flat space line element $ds^{2}_{2D} =
-dr^{2} + r^{2} d\theta^{2}$ and the aforementioned integral of motion
$Q=p_{\theta}$. At the quantum level we can thus write the eigenvalue
equation
\begin{equation}
\hat{Q} \Psi(r,\theta) = \kappa\Psi(r,\theta) \Rightarrow-\mathrm{i}
\hbar\frac{\partial\Psi}{\partial\theta} = \kappa\Psi,
\end{equation}
which leads to the solution $\Psi(r,\theta) = \frac{1}{\sqrt{2\pi\hbar}}
e^{\frac{\mathrm{i}}{\hbar}\kappa\theta} \psi(r)$. Note that here we have no reason to consider $\theta$ as a periodic variable. As a result we take $\kappa$ to have a continuous spectrum and be normalized to a Dirac delta function, like the eigenvalues $\mu,\nu$ previously. The $\psi(r)$ part is to be
obtained by the Wheeler-DeWitt equation which results to
\begin{equation}%
\begin{split}
\widehat{\mathcal{H}}\Psi(r,\theta)=0 \Rightarrow &  \left[  -h^{2} \left(
-\frac{\partial^{2}}{\partial r^{2}}-\frac{1}{r}\frac{\partial}{\partial r}
+\frac{1}{r^{2}}\frac{\partial^{2}}{\partial\theta^{2}}\right)  +\mathcal{V}%
_{0}^{2}\right]  \Psi(r,\theta) =0 \Rightarrow\\
&  \frac{1}{r} \frac{d}{d r}\left(  r\frac{d}{d r} \psi(r) \right)  + \left(
\frac{\mathcal{V}_{0}^{2}}{\hbar^{2}} +\frac{\kappa^{2}}{\hbar^{2} r^{2}}
\right)  \psi(r) =0 .
\end{split}
\end{equation}
The latter is the Bessel equation with general solution
\begin{equation}
\psi(r) = C_{1} J_{\frac{\mathrm{i} \kappa}{\hbar}}\left(  \frac
{\mathcal{V}_{0}}{\hbar} r\right)  + C_{2} Y_{\frac{\mathrm{i} \kappa}{\hbar}%
}\left(  \frac{\mathcal{V}_{0}}{\hbar} r\right)  ,
\end{equation}
where the $J_{n}(z)$, $Y_{n}(z)$ are the Bessel equations of the first and
second kind respectively.

In \cite{Gryb,Gielen}, where a similar Wheeler-DeWitt equation is explored, a certain linear combination of the solution is chosen which leads to a delta function normalization. In particular if you take as wave function
\begin{equation} \label{choice1}
  \psi (r) \propto
\mathrm{Re}\left[ e^{-\mathrm{i}\ln \sigma }  J_{\mathrm{i} s}(\sigma r)\right],
\end{equation}
where $\sigma = \frac{\mathcal{V}_{0}}{\hbar}$ and $s=\frac{\kappa}{\hbar}$ it can be seen that \cite{Gryb,Gielen}
\begin{equation} \label{orthoth}
  \int_{0}^{+\infty}\!\! r|\psi(r)^2| dr \propto \delta(\sigma-\sigma') .
\end{equation}
Note that the weight $r$ in the integral in \eqref{orthoth} is exactly what
emerges from using the natural measure for the inner product between states,
i.e. the square root of the determinant of the mini-superspace metric (in the
$(r,\theta)$ coordinates $|\det(\bar{G}_{IJ})|^{1/2}=r$).
The additional phase $e^{-\mathrm{i}\ln \sigma }$ in \eqref{choice1} is introduced in order to eliminate finite terms appearing to the lower limit of the above integral.

However here we explore another linear combination in which we use as our base the function defined firstly in \cite{Dunster} and denoted by
\begin{equation}
W_{s,\sigma}(r) = \frac{1}{\cosh\left(  \frac{\pi s}{2}\right)  }
\mathrm{Re}\left[  J_{\mathrm{i} s}(\sigma r)\right] .
\end{equation}
It forms a different linear combination of the solution and it is shown to be still normalizable in terms of a delta function \cite{NiAn}. Unlike \eqref{choice1} it lacks the phase which serves to eliminate the $\sigma$ dependence of the wave function at the limit $r\rightarrow 0$.
By using the typical procedure of
deriving the orthogonality condition in a Sturm-Liouville problem, it has been
shown that (see the appendix of \cite{NiAn})
\begin{equation}
\label{Worth}\int_{0}^{+\infty}\!\! r W_{s,\sigma}(r)W_{s,\sigma^{\prime}}(r)
dr =\frac{1}{\sigma} \delta(\sigma- \sigma^{\prime}) + \frac{\mathcal{A}_0 (\sigma,\sigma')}{\sigma^2- \sigma'^2}
\end{equation}
The additional term $\mathcal{A}_0$ is given by
\begin{equation}
  \mathcal{A}_0 = \frac{2}{\pi} \tanh\left( \frac{\pi s}{2} \right) \sin\left[ s \ln \left(\frac{\sigma}{\sigma'} \right)  \right]
\end{equation}
and it is eliminated under the condition
\begin{equation}
\label{orthocond}\ln\left(  \frac{\sigma}{\sigma^{\prime}}\right)  = \frac{k
\pi}{s}, \quad k \in\mathbb{Z} .
\end{equation}
Thus a quantum restriction is set upon $\sigma$ to guarantee orthogonality when $\sigma\neq \sigma'$. In the case $\sigma=\sigma'$ the limit
\begin{equation}
  \lim_{\sigma\rightarrow \sigma'} \left(\frac{\mathcal{A}_0 (\sigma,\sigma')}{\sigma^2- \sigma'^2} \right) = s \frac{\tanh \left( \frac{\pi s}{2} \right)}{\pi \sigma^2}
\end{equation}
is finite and just adds a constant to the delta function.
The necessary for the orthogonality of states condition \eqref{orthocond}, ensures
at the same time the Hermiticity of the operator $\hat{\mathcal{H}}$ under the
assumption that the wave function vanishes at the boundary of the half line
$(0,+\infty)$.

As a result we may write the full wave function of this case as
\begin{equation}
\label{wave2}\Psi_{\kappa,\mathcal{V}_{0}}(r,\theta) = \sqrt{\frac
{\mathcal{V}_{0} \hbar}{2\pi}} \frac{e^{\frac{\mathrm{i}}{\hbar}\kappa\theta}%
}{\cosh\left(  \frac{\pi\kappa}{2\hbar}\right)  } \mathrm{Re}\left[
J_{\frac{\mathrm{i} \kappa}{\hbar}}\left(  \frac{\mathcal{V}_{0}}{\hbar}
r\right)  \right]  ,
\end{equation}
satisfying
\begin{equation}
\label{w2orth}\int_{-\infty}^{+\infty}\int_{0}^{+\infty} r \Psi_{\kappa
^{\prime},\mathcal{V}_{0}^{\prime}}^{*} \Psi_{\kappa,\mathcal{V}_{0}} d\theta
dr = \delta(\kappa-\kappa^{\prime}) \delta(\mathcal{V}_{0}-\mathcal{V}%
_{0}^{\prime})
\end{equation}
subject to the condition
\begin{equation}
\label{orthocondV}\ln\left(  \frac{\mathcal{V}_{0}}{\mathcal{V}_{0}^{\prime}%
}\right)  = \frac{k \pi}{\kappa} \hbar, \quad k \in\mathbb{Z}.
\end{equation}
As it is evident from \eqref{wave2}-\eqref{orthocondV}, we choose to interpret
$\mathcal{V}_{0}$ as some short of ``eigenvalue". In this manner we observe
that for a fixed $\kappa$, a discretization is introduced in the fiducial
volume of the three space through the orthogonality condition
\eqref{orthocondV}. It is interesting that the study of the reduced
mini-superspace system can yield such an information about the three space,
which is usually discarded  through the process of the reduction.

\section{Classical solution}

\label{sec6}

We lastly proceed with the presentation of the classical solution for the
gravitational field equations for arbitrary potential . For the sake of
simplicity of the resulting expressions we here present the case where
$B\left(  \phi\right)  =\frac{B_{1}}{V\left(  \phi\right)  }$. For the lapse
parameterization $N\left(  t\right)  =\bar{N}\left(  t\right)  \left(
a^{3}V\left(  \phi\right)  \right)  ^{-1}$ (in which the potential of the
relevant Lagrangian \eqref{ac.05} is free of $a, \phi$ )  the aforementioned
equations become%
\begin{equation}
\frac{1}{\bar{N}^{2}}\left(  3B_{1}a^{4}\dot{a}^{2}-\frac{a^{6}}{2}\dot{\psi}%
^{2}\right)  -1=0, \label{sr.01}%
\end{equation}%
\begin{equation}
6B_{1}a^{4}\ddot{a}+12B_{1}a^{3}\dot{a}^{2}+3a^{5}\dot{\psi}^{2}-\frac{6}%
{\bar{N}}B_{1}a^{4}\left(  \bar{N}\right)  ^{\cdot}\dot{a}=0, \label{sr.02}%
\end{equation}%
\begin{equation}
\ddot{\psi}+6\frac{\dot{a}}{a}\dot{\psi}-\frac{\bar{N}}{\bar{N}}\dot{\psi}=0.
\label{sr.03}%
\end{equation}
where the new field $\psi$ is defined as $d\psi=\sqrt{V\left(  \phi\right)
}d\phi$.

For $\bar{N}\left(  t\right)  =1$, these equations can be integrated resulting
in the scale factor
\begin{equation}
a\left(  t\right)^{6}=\frac{3}{B_{1}}t^{2}+\alpha_{1}%
\frac{\sqrt{6}}{\sqrt{B_{1}}}t+ \frac{\alpha_{2}}{2}%
\end{equation}
and the scalar field $\psi\left(  t\right)  $
\begin{equation}
\psi\left(  t\right)  =\sqrt{\frac{2B_1}{3}}\mathrm{arctanh}\left(  \frac{\sqrt{6}t+\sqrt{B_{1}%
}\alpha_{1}}{\sqrt{B_{1}\left(a_{1}^{2}-\alpha_2\right)  }}\right)  ,
\end{equation}
while the line element (\ref{ac.03}) reads%
\begin{equation}
ds^{2}=-\frac{1}{a^{6}\left(  t\right)  V\left(  \phi\left(  t\right)
\right)  }dt^{2}+a^{2}\left(  t\right)  \left(  dx^{2}+dy^{2}+dz^{2}\right)  .
\end{equation}

We observe that the potential functions $V\left(  \phi\right)  $ has been
included in the line element and consequently it affects all the geometric and
physical quantities. The solution that we constructed here is known as
algebraic solution because the physical quantities are given by algebraic
equations.\ This kind of solution has been before derived for the quintessence
field in \cite{ns1} with various physical applications \cite{ns2,ns3}.

\section{Conclusions}

\label{sec7}

In this piece of work we considered an Einstein-aether scalar field
cosmological theory proposed as a Lorentz violating inflationary model. The
scalar and aether fields are interacting due to the assumption that the
constants of the aether part of the action are taken to be functions of the
scalar field. A useful and critical property of this theory is that, for the
assumed geometry and the consequent assumptions for the fields, the reduced
field equations are correctly described by the corresponding minisuperspace
Lagrangian inferred by the reduced action. This occurrence is not at all
automatic for arbitrary reductions, and is certainly not common in
Einstein-aether theories. Yet, it is an important facilitation in order to
study the quantization process.

For the spatially flat FLRW geometry considered, the metric of the
two-dimensional minisuperspace (spanned by $a,\phi$) depends on two unknown
functions, the scalar field potential $V\left(  \phi\right)  $ and the
collective gravitational coupling function for the aether field, $B\left(
\phi\right)  $. For a specific relation of the two unknown functions, for
which the configuration manifold becomes flat, we were able to: (A) write the
general algebraic classical solution to the simplified cosmological field
equations; and (B) present the quantization of the model which is carried out
in the flat coordinates of the configuration space.

Surprisingly enough, the Wheeler-DeWitt equation is revealed as that of a free particle in a two dimensional flat space of hyperbolic signature. Under an appropriate choice for the basic wave function, certain quantization conditions can be enforced in the constant appearing due to the spatial
integration at the classical level and which represents the spatial volume of the three space.

We note that the above quantization procedure encompasses an infinitely large class of models since it is based in a combination relating the function $B(\phi)$ with the potential $V(\phi)$ (see \eqref{lawV} or \eqref{lawB}). Even a specific $B(\phi)$ by itself contains infinitely many possible combinations of the $\beta_i(\phi)$ of the original Lagrangian that can realize it. As a result, the physical implications of this process of quantization are expected to have different interpretations in each specific model that belongs to the class characterized by \eqref{lawV} and \eqref{lawB}.

In a future work we plan to apply the classical solution in order to study the
physical applications of the model. Also, to investigate the general case of
unrelated $V\left(  \phi\right) $, $B\left(  \phi\right) $ as well as
different geometries, such as Bianchi $I, V$.

\begin{acknowledgments}
AP \& GL were funded by Agencia Nacional de Investigaci\'{o}n y Desarrollo -
ANID through the program FONDECYT Iniciaci\'{o}n grant no. 11180126.
Additionally, GL is supported by Vicerrector\'{\i}a de Investigaci\'{o}n y
Desarrollo Tecnol\'{o}gico at Universidad Cat\'olica del Norte.
\end{acknowledgments}

\end{document}